\documentclass[showpacs,aps,prd,floatfix,amsmath,amssymb,nofootinbib,superscriptaddress]{revtex4-2}
\usepackage{graphicx}
\usepackage{epstopdf}
\usepackage{amssymb}
\usepackage{bbm}
\usepackage[utf8]{inputenc}
\usepackage[dvipsnames]{xcolor}
\usepackage{newunicodechar}
\newunicodechar{−}{-}
\usepackage{slashed}
\usepackage{float}
\usepackage{subfigure}
\usepackage{hyperref}
\begin{document}

\title{Forward--backward asymmetry in the $H\to Z\gamma\to\overline{f}{f}\gamma $ decay}
\author{A.I. Hern\'andez-Ju\'arez}
\email{alan.hernandezjua@alumno.buap.mx}
\address{Facultad de Ciencias F\'isico Matem\'aticas, Benem\'erita Universidad Aut\'onoma de Puebla, Apartado Postal 1152, Puebla, Pue., M\'exico.}
\author{D. Watko}
\address{Facultad de Ciencias F\'isico Matem\'aticas, Benem\'erita Universidad Aut\'onoma de Puebla, Apartado Postal 1152, Puebla, Pue., M\'exico.}
\author{A. Fern\'andez-T\'ellez}
\address{Facultad de Ciencias F\'isico Matem\'aticas, Benem\'erita Universidad Aut\'onoma de Puebla, Apartado Postal 1152, Puebla, Pue., M\'exico.}
\author{G. Tavares-Velasco}
\address{Facultad de Ciencias F\'isico Matem\'aticas, Benem\'erita Universidad Aut\'onoma de Puebla, Apartado Postal 1152, Puebla, Pue., M\'exico.}

\date{\today}

\begin{abstract}
We study the forward--backward asymmetry in the three-body decay process $H\to Z\gamma\to\overline{f}f\gamma$, induced by complex form factors in the $HZ\gamma$ vertex. To estimate its magnitude, we derive constraints on the real and imaginary parts of the $CP$-violating form factor $h_3^{Z\gamma}$ using current LHC measurements, obtaining upper limits of about $0.9$ GeV. These bounds are of the same order of magnitude as those derived from electric dipole moments (EDM). We find that the asymmetry can reach values of order $10^{-1}$ and may be accessible at the HL-LHC.
\end{abstract}


\maketitle

\section{Introduction}

The $H\to Z\gamma$ decay attracted considerable attention
following the first experimental evidence reported by the ATLAS and CMS collaborations, with a measured signal strength of $\mu^{Z\gamma}=2.2\pm 0.7$ \cite{CMS:2022ahq, ATLAS:2023yqk}. Although this result was consistent with the Standard Model (SM) within $1.9\sigma$, it was regarded as a potential window into new physics. In particular, the central value suggested a possible enhancement with respect to the SM prediction, motivating further theoretical and phenomenological studies. Several works explored possible explanations for this deviation within extensions of the SM, including two-Higgs-doublet models
\cite{Chen:2024vyn, Sang:2024vqk, Chen:2024oru},
non-supersymmetric scenarios \cite{Benbrik:2022bol},
the minimal supersymmetric SM \cite{Israr:2024ubp, ReyesR:2025dok},
and models with additional fermions or gauge bosons
\cite{Barducci:2023zml, Lichtenstein:2023vza, Boto:2023bpg, He:2024bxi, Cheung:2024kml, Kachanovich:2025cxz}.
In this context, $CP$ violation in the $HZ\gamma$ vertex
was also proposed to explain the deviations in
$\mu^{Z\gamma}$ reported by the LHC \cite{Hernandez-Juarez:2024iwe}. These contributions arise from flavor-changing neutral currents (FCNC) involving the $Z$ and Higgs boson.
 
\begin{figure}[H]
\begin{center}
\includegraphics[width=9cm]{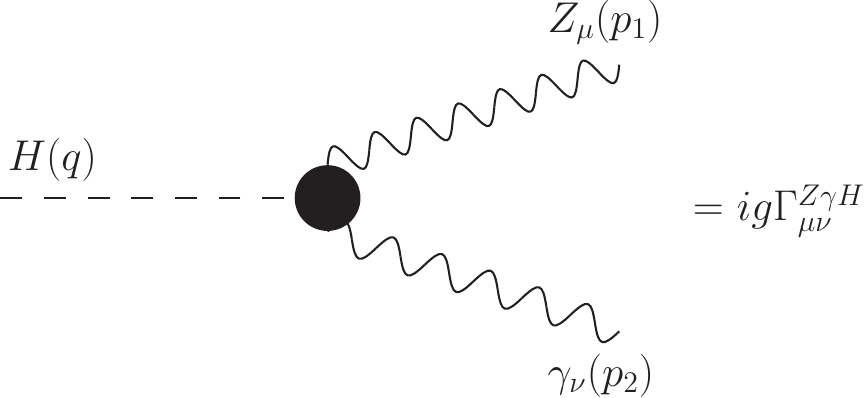}
\caption{The $HZ\gamma$ vertex function. The Feynman diagrams in this work are created with \texttt{JaxoDraw} \cite{Binosi:2003yf}.} \label{HWWvertex}
\end{center}
\end{figure}

The $CP$-even and $CP$-odd form factors can be introduced in the most general $HZ\gamma$ effective vertex as follows 
\begin{equation}
\label{VertexFunction}
\Gamma_{Z\gamma H}^{\mu\nu}=h_1^{Z\gamma} \Bigg\{ g^{\mu\nu}+\frac{2}{m_Z^2-m_H^2} p_1^\nu p_2^\mu\Bigg\}+\frac{1}{m_Z^2}h^{Z\gamma}_3\epsilon^{\mu\nu\alpha\beta}p_{1\alpha}p_{2\beta},
\end{equation}
where we adopt the kinematics shown in Fig. \ref{HWWvertex}. The form factors $h_1^{Z\gamma}$ and $h_3^{Z\gamma}$ are $CP$-conserving and $CP$-violating, respectively. The notation in Eq. \ref{VertexFunction} implies that the form factors have units of mass. Furthermore, gauge invariance relates $h_1^{Z\gamma}$ to $h_2^{Z\gamma}$, and therefore the latter is not considered. In the SM, only $h_1^{Z\gamma}$ is generated at the one-loop level \cite{Cahn:1978nz, Bergstrom:1985hp, Gunion:1989we, DECKER1991605}, with dominant contributions arising from $W$ boson loops. Fermion loops also contribute, giving rise to a small imaginary part. The total electroweak contribution is given by
\begin{equation}
    h_1^{Z\gamma}=(-3.22\times 10^{-1}+i\, 2.6\times 10^{-4})\text{ GeV}.\label{h1num}
\end{equation}
QCD corrections have been computed and are known to be small
\cite{SPIRA1992350,Bonciani:2015eua, Gehrmann:2015dua}, whereas next-to-leading-order electroweak corrections can reach $\sim 7\%$ of the leading contribution \cite{Sang:2024vqk}. Imaginary parts such as that in Eq. \ref{h1num}  also arise in other SM radiative corrections  \cite{Hernandez-Juarez:2020drn,Gounaris:2000tb,Hernandez-Juarez:2023dor} and can play an important role
in $CP$-violating observables \cite{Godbole:2007cn,Hernandez-Juarez:2021xhy, Hernandez-Juarez:2025nzh}.  The study of $CP$-odd effects in the $HZ\gamma$ coupling
has been performed in $\gamma p$ \cite{TaheriMonfared:2016gua}
and $e^+e^-$ colliders \cite{Hagiwara:2000tk,Rindani:2009pb},
as well as through polarized observables
\cite{Korchin:2013ifa, Ahmed:2023vyl, Hasan:2024rib}. Further studies can be found in Refs. \cite{Hankele:2006ma, Korchin:2013jja, He:2020suf}.

The general conditions for generating $CP$-violating observables in the $HZ\gamma$ coupling with three reconstructable final-state particles were discussed in Ref.~\cite{Chen:2014ona}. In this framework, such observables arise from the interference between weak phases in the Higgs Lagrangian and strong phases induced by the Breit--Wigner widths of different intermediate particles. Similar interference mechanisms have been studied in $B$-meson physics \cite{Eilam:1991yv, Atwood:1994zm, Bediaga:2009tr}, top-quark decays \cite{Eilam:1991yv, NOWAKOWSKI_1991}, and supersymmetric models \cite{Berger:2011wh}.

In the particular case of the $HZ\gamma$ coupling, $CP$-violating effects can be probed through a forward--backward asymmetry ($\mathcal{A}_{FB}$). Such an asymmetry can arise either from the interference between $CP$-even and $CP$-odd complex form factors in the $HZ\gamma$ vertex \cite{Gao:2010qx, Bolognesi:2012mm, Anderson:2013afp}, or from the interference of the signal with intermediate states or background processes \cite{Kachanovich:2020xyg}, such as $H\to \gamma^\ast\gamma \to \overline{\ell}\ell\gamma$ \cite{Chen:2014ona}, $H\to ZZ\to 4\ell$ \cite{He:2019kgh}, $gg\to ZZ\to 4\ell$ \cite{Feng:2021izk}, $gg\to \gamma^\ast\gamma \to \overline{\ell}\ell\gamma$ \cite{Korchin:2014kha}, and $gg\to Z\gamma\to \overline{\ell}\ell\gamma$ \cite{Chen:2017plj}. However, the former mechanism, which originates solely from the $HZ\gamma$ vertex, has received comparatively less attention in recent studies \cite{Gritsan:2022php}.

In this work, we focus on the forward--backward asymmetry
induced by complex $CP$-conserving and $CP$-violating form factors in the $HZ\gamma$ vertex.
We show that this asymmetry can be generated without the need for interference with background processes. Furthermore, bounds on the $CP$-violating form factor $h_3^{Z\gamma}$
are obtained from current LHC data and used to evaluate the size
of the forward--backward asymmetry. The manuscript is organized as follows: in Sec \ref{secCP}, we compute the $H\to Z\gamma\to\overline{f}f\gamma$ decay amplitude and derive the forward--backward asymmetry. In Sec. \ref{secbounds}, limits on the $CP$-violating form factor $h_3^{Z\gamma}$ are obtained, whereas in Sec. \ref{numsec} we present a numerical analysis of the
$CP$-violating effects and the possibility of observing $\mathcal{A}_{FB}$ in the high-luminosity LHC (HL-LHC).
Finally, in Sec.~\ref{secconclu}, we present our conclusions.

\section{$CP$ violation in the $H\to Z\gamma\to\overline{f}f\gamma$ decay}\label{secCP}

In this section, we analyze the effects of the $CP$-violating form factor $h_3^{Z\gamma}$ on the decay $H\to Z\gamma\to\overline{f}f\gamma$. Since the form factor $h_1^{Z\gamma}$ is complex within the SM, we allow $h_3^{Z\gamma}$ to be complex as well. To identify the role of the absorptive parts, it is convenient to write $h_i^{Z\gamma}$ ($i=$1, 3) as
\begin{equation}
   h_i^{Z\gamma} = {\rm Re}\big[h_i^{Z\gamma}\big] + i\, {\rm Im}\big[h_i^{Z\gamma}\big],\quad i=1\text{, }3.
\end{equation}
In the SM, complex $CP$-conserving form factors have been studied in scenarios where at least one of the particles in the vertex is off-shell \cite{Hernandez-Juarez:2020drn,Gounaris:2000tb,Hernandez-Juarez:2023dor}. In contrast, $CP$-violating anomalous couplings with nonzero imaginary parts can appear  in extensions of the SM, such as in $ZZV^\ast$ and $Z\gamma V^\ast$ ($V=Z,\gamma$) vertices \cite{Hernandez-Juarez:2021mhi, Hernandez-Juarez:2022kjx}, as well as in quark-gluon interactions \cite{Hernandez-Juarez:2020gxp, Aranda:2018zis, Aranda:2020tox}.

\begin{figure}[H]
\begin{center}
\includegraphics[width=7cm]{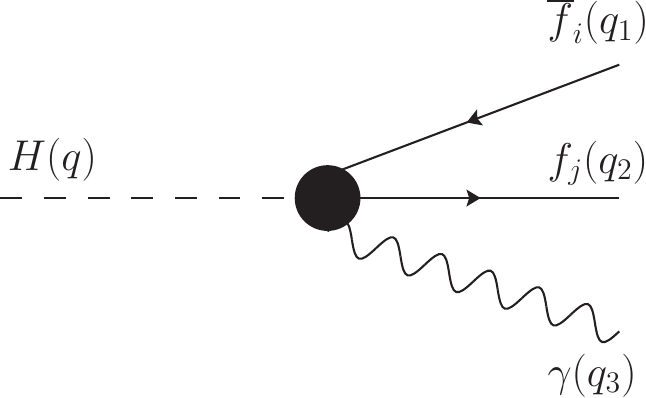}
\caption{Kinematic configuration and momentum assignments for the $HZ\gamma$ vertex.} \label{decaynom}
\end{center}
\end{figure}

We follow the momentum assignment shown in Fig.~\ref{decaynom} and perform the calculations with \texttt{FeynCalc} \cite{Mertig:1990an, Shtabovenko:2016sxi, Shtabovenko:2020gxv, Shtabovenko:2023idz}. The process $H \to Z\gamma \to \bar{f} f \gamma$ proceeds through an intermediate $Z$ boson decaying into a fermion pair. The corresponding amplitude can thus be written as the product of the $H \to Z\gamma$ and $Z \to \bar{f} f$ amplitudes, linked by the $Z$-boson propagator:
\begin{equation}
\mathcal{M}(H\to Z\gamma\to \bar{f} f \gamma)
=
-i\,\frac{g
}
{k^2 - m_Z^2 + i\,\Gamma_Z m_Z}\mathcal{M}^{\mu\nu}(H\to Z(k)\gamma)
\,
\mathcal{M}_{\mu}(Z(k)\to \bar{f} f)\epsilon_\nu^{\ast}(p_3,\lambda)
\,.
\end{equation}
Since the fermions are treated as massless, the term $k^\sigma k^\mu/m_Z^2$ in the $Z$ boson propagator does not contribute to the squared amplitude and can therefore be safely neglected. Using Eq.~\eqref{VertexFunction}, the $H \to Z\gamma$ decay amplitude, including complex form factors, can be written as
\begin{align}
   \mathcal{M}^{\mu\nu}(H\to Z(k)\gamma) = ig \Bigg[ \Big({\rm Re}\big[h_1^{Z\gamma}\big] + i\, {\rm Im}\big[h_1^{Z\gamma}\big]\Big) \Bigg\{ g^{\mu\nu}+\frac{2}{m_Z^2-m_H^2} k^\nu q_3^\mu\Bigg\}+\frac{1}{m_Z^2}\Big({\rm Re}\big[h_3^{Z\gamma}\big] + i\, {\rm Im}\big[h_3^{Z\gamma}\big]\Big)\epsilon^{\mu\nu\alpha\beta}k_{\alpha}q_{3\beta}\Bigg],
\end{align}
with $k=q_1+q_2$. For the $Z$ boson decay to fermions, the amplitude is
\begin{align}
    \mathcal{M}^{\mu}(Z(k)\to \bar{f} f)=i\frac{g}{c_W}\overline{u}(q_1)\gamma^\mu(g_V-\gamma_5g_A)u(q_2).
\end{align}
Following the phase-space parametrization detailed in Appendix~\ref{spacephaseapp}, the differential partial width can be expressed as\begin{equation}\label{dif1width}
\frac{d\Gamma}{dK\, d\cos\theta}=\frac{m_H^2-m_Z^2}{512\, \pi^3 m_H^3}\mathcal{M}^2(H\to Z\gamma\to\overline{f}f\gamma), 
\end{equation}
where $K \equiv k^2$ denotes the invariant mass of the $Z$ boson. The angle $\theta$ is defined between the antifermion three-momentum $\vec{q}_1$ and the $x$-axis in the $Z$ rest frame, as shown in Fig.~\ref{plano}. The corresponding squared amplitude is given by
\begin{align}
 \mathcal{M}^2=&   
\frac{
4\, g^4\, N_c 
}{
c_W^2 m_Z^2
\Big(
(k^2-m_Z^2)^2+m_Z^2\Gamma_Z^2
\Big)
}
\Bigg[
(g_A^2+g_V^2)\,|h_3^{Z\gamma}|^2\,(q_1\cdot q_3)^2\nonumber\\
&
-4\,(q_1\cdot q_3)
\Bigg(
g_A g_V\,{\rm Im}\!\left[h_1^{Z\gamma}(h_3^{Z\gamma})^*\right]\,m_Z^2
+2(g_A^2+g_V^2)\,|h_1^{Z\gamma}|^2\,
\frac{m_Z^4\,(q_2\cdot q_3)}{(m_H^2-m_Z^2)^2}
\Bigg)\nonumber\\
&
+(g_A^2+g_V^2)\,|h_1^{Z\gamma}|^2\,m_Z^4
+4 g_A g_V\,{\rm Im}\!\left[h_1^{Z\gamma}(h_3^{Z\gamma})^*\right]\,m_Z^2\, (q_2\cdot q_3)
+(g_A^2+g_V^2)\,|h_3^{Z\gamma}|^2\,(q_2\cdot q_3)^2.
\Bigg],\label{squaredamp}
\end{align}
where $N_c$ corresponds to the number of colors ($N_c=1$ for leptons). The squared amplitude depends on the quadratic terms $|h_i^{Z\gamma}|^2$ and the interference term ${\rm Im}\!\left[h_1^{Z\gamma}(h_3^{Z\gamma})^*\right]$, which are defined as 
\begin{equation}
|h_i^{Z\gamma}|^2
= {\rm Re}\big[h_i^{Z\gamma}\big]^2 + {\rm Im}\big[h_i^{Z\gamma}\big]^2,
\end{equation}
\begin{equation}
    {\rm Im}\!\left[h_1^{Z\gamma}(h_3^{Z\gamma})^*\right]
=
{\rm Im}\big[h_1^{Z\gamma}\big]\,{\rm Re}\big[h_3^{Z\gamma}\big]
-
{\rm Re}\big[h_1^{Z\gamma}\big]\,{\rm Im}\big[h_3^{Z\gamma}\big].
\end{equation}
 Additionally, the $\mathcal{M}^2$ is expressed in terms of the Lorentz invariants $q_i\cdot q_3$ ($i=1,2$), whose explicit forms in the Higgs rest frame are given in Appendix~\ref{lorinvcal}. Since the $Z$ boson produced in the decay is on-shell, we employ the narrow-width approximation:
\begin{equation}
    \lim_{m_Z\Gamma_Z\rightarrow0}\frac{1}{(k^2-m_Z^2)^2+m_Z^2\Gamma_Z^2}=\delta(k^2-m_Z^2)\frac{\pi}{m_Z\Gamma_Z},
\end{equation}
which allows the analytic integration over the invariant mass
$K=k^2$. The differential partial width can be expressed as follows: 
\begin{align}
 \frac{d\Gamma(H\to Z\gamma\to \overline{f}f\gamma)}{d\cos\theta} =&
\frac{g^4\,(m_H^2-m_Z^2)\, N_c}{1024\,c_W^2\,m_H^3\,m_Z^3\,\pi^2\,\Gamma_Z}\Bigg[
-16\,g_A g_V\,{\rm Im}\!\left[h_1^{Z\gamma}(h_3^{Z\gamma})^*\right]
\,m_Z^2\,(m_H^2-m_Z^2)\,\cos\theta \nonumber\\
&+\,(g_A^2+g_V^2)\Big\{
4|h_1^{Z\gamma}|^2
m_Z^4
+|h_3^{Z\gamma}|^2
\,(m_H^2-m_Z^2)^2
\Big\}(1+\cos^2\theta)\Bigg].\label{difpart} 
\end{align}
The term proportional to interference
${\rm Im}\!\left[h_1^{Z\gamma}(h_3^{Z\gamma})^*\right]$
is odd under $\cos\theta\to-\cos\theta$ and therefore
generates a forward--backward asymmetry in the angular distribution, which provides a signal of $CP$ violation. The structure of the differential partial width in Eq.~\eqref{difpart} is similar to that found in Ref.~\cite{Chen:2014ona}. In that case, the corresponding width is generated by the interference between the $H\to \gamma^\ast\gamma$ and $H\to Z\gamma$ intermediate processes, which contribute to the same three-body final state.

Integrating Eq.~\eqref{difpart} yields
\begin{align}
\Gamma(H\to Z\gamma\to \overline{f}f\gamma) =   
\frac{
g^4\,(g_A^2+g_V^2)\,(m_H^2-m_Z^2)\, N_c}{
384\,c_W^2\,m_H^3\,m_Z^3\,\pi^2\,\Gamma_Z
}
\Big[
4|h_1^{Z\gamma}|^2
m_Z^4
+|h_3^{Z\gamma}|^2
(m_H^2-m_Z^2)^2
\Big]
,\label{finalwidth}
\end{align}
which can be expressed in terms of the partial width and branching ratio of the $H\to Z\gamma$ and $Z\to \overline{f}f$ decays, respectively. These quantities are given by
\begin{equation}
\Gamma(H\to Z\gamma)=g^2\frac{m_H^2-m_Z^2}{32\pi\, m_H^3\, m_Z^4}\Big\lbrace 4|h_1^{Z\gamma}|^2 m_Z^4 +|h_3^{Z\gamma}|^2(m_H^2-m_Z^2)^2 \Big\rbrace
\end{equation}
\begin{equation}
\mathcal{B}(Z\to\overline{f}f)=\frac{g^2 m_Z\, N_c}{12 \pi\, c^2_W \Gamma_Z}\big(g_A^2+g_V^2\big).
\end{equation}
The partial width in Eq.~\eqref{finalwidth} depends quadratically on $h_3^{Z\gamma}$, whereas the angular distribution in Eq.~\eqref{difpart} includes linear terms. As a result, angular observables exhibit enhanced sensitivity to the $CP$-odd form factor. We therefore concentrate on the forward--backward asymmetry.

\subsection{Forward--backward asymmetry}

The $CP$-violating effects in the $HZ\gamma$ vertex can be probed through
the forward--backward asymmetry ($\mathcal{A}_{FB}$) in the angular distribution of the final-state fermions. This asymmetry may arise from interference between the signal and background processes
\cite{Chen:2017plj,Feng:2021izk,Chen:2014ona, Korchin:2014kha}. However, as indicated by Eq.~\eqref{difpart}, a nonvanishing $\mathcal{A}_{FB}$ can also be generated directly from the decay $H\to Z\gamma\to \overline{f}f\gamma$, even in the absence of background contributions. The forward--backward asymmetry is defined as
\begin{align}
    \mathcal{A}_{FB}&=\frac{\int _0^1d\cos\theta\frac{d\Gamma(H\to Z\gamma\to\overline{f}f\gamma)}{d\cos\theta}-\int _{-1}^0d\cos\theta\frac{d\Gamma(H\to Z\gamma \to\overline{f}f\gamma)}{d\cos\theta}}{\int _0^1d\cos\theta\frac{d\Gamma(H\to Z\gamma\to\overline{f}f\gamma)}{d\cos\theta}+\int _{-1}^0d\cos\theta\frac{d\Gamma(H\to Z\gamma\to\overline{f}f\gamma)}{d\cos\theta}},\label{afb1}
\end{align}
Although this asymmetry has been discussed in the literature \cite{Gao:2010qx, Bolognesi:2012mm, Anderson:2013afp}, to the best of our knowledge, it has not been presented explicitly in analytic form. After performing the angular integration in Eq. \eqref{afb1}, we obtain
\begin{align}
    \mathcal{A}_{FB}
    &=\frac{6\, g_A g_V m_Z^{2}(m_H^{2}-m_Z^{2})
\, N_c}{(g_A^{2}+g_V^{2})}\Bigg\{\frac{-
{\rm Im}\!\left[h_1^{Z\gamma}(h_3^{Z\gamma})^*\right]
}{
4|h_1^{Z\gamma}|^2\,m_Z^{4}
+|h_3^{Z\gamma}|^2(m_H^{2}-m_Z^{2})^{2}
}\Bigg\}.\label{FW-exp}
\end{align}
From this expression, it follows that a non-vanishing $\mathcal{A}_{FB}$ requires $CP$-violating and complex form factors. Since $h_1^{Z\gamma}$ is complex in the SM, a nonzero $h_3^{Z\gamma}$ is sufficient to induce the forward--backward asymmetry.  The magnitude of $\mathcal{A}_{FB}$ is controlled by the real and absorptive parts of $h_3^{Z\gamma}$. Consequently, estimating the size of the asymmetry requires an exploration of the allowed parameter space of the $CP$-violating form factor.


\section{Bounds on $h_3^{Z\gamma}$} \label{secbounds}

\begin{figure}[H]
\begin{center}
\includegraphics[width=9cm]{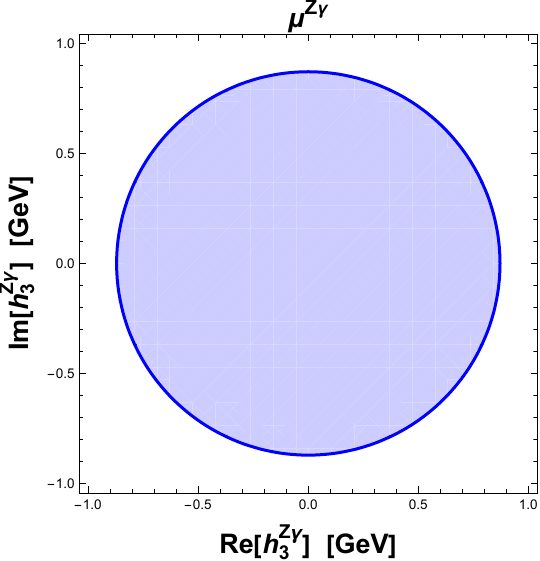}
\caption{Allowed region for the real and imaginary parts of the $CP$-violating form factor
$h_3^{Z\gamma}$ at the 95\% CL, derived from the signal strength
$\mu^{Z\gamma}=1.3^{+0.6}_{-0.5}$ reported by the ATLAS collaboration
\cite{ATLAS:2025aip}.} \label{boundn}
\end{center}
\end{figure}

In this section, we obtain constraints on the real and imaginary parts of the $CP$-violating form factor $h_3^{Z\gamma}$ to evaluate the $\mathcal{A}_{FB}$ asymmetry. In Ref. \cite{Hernandez-Juarez:2024iwe}, it was shown that, in the absence of new-physics effects in Higgs production, the signal strength of the $gg\to H\to Z\gamma$ decay can be approximated as 
\begin{align}
\label{WidthGeneral}
\mu^{Z\gamma}&\approx\frac{\mathcal{B}^{SM}(H\to Z\gamma)+\delta\Gamma^{CP}(H\to Z\gamma)/\Gamma_H}{\mathcal{B}^{SM}(H\to Z\gamma)}
\end{align}
where $\mathcal{B}^{SM}(H\to Z\gamma)$ and $\Gamma_H$ denote the SM branching ratio for the $H\to Z\gamma$ decay and the total width of the Higgs boson, respectively. The correction  $\delta\Gamma^{CP}$ arises from the $CP$-odd form factor $h_3^{Z\gamma}$, and is given by
\begin{equation}\label{delta}
\delta\Gamma^{CP}(H\rightarrow Z\gamma)= g^2\frac{\big(m_H^2-m_Z^2\big)^3 }{32\ \pi m_H^3 m_Z^4} |h^{Z\gamma}_3|^2.
\end{equation}
Using $\mathcal{B}^{SM}(H\to Z\gamma)=1.57\times10^{-3}$ and $\Gamma_H= 4.1$ MeV \cite{LHCHiggsCrossSectionWorkingGroup:2016ypw},  Eq. \eqref{WidthGeneral} can be expressed in terms of the real and imaginary parts of $h_3^{Z\gamma}$ as follows
 \begin{align}
\label{Ratio2}
\mu^{Z\gamma}&=1+(1.94 \text{ GeV}^{-2}) \Big({\rm Re}\big[h^{Z\gamma}_3\big]^2+{\rm Im}\big[h^{Z\gamma}_3\big]^2\Big),
\end{align}
where the QCD NLO contributions are neglected, as they are expected to be small \cite{SPIRA1992350}. Previous limits on $h_3^{Z\gamma}$ have been obtained from $e^+e^-$ colliders \cite{Hagiwara:2000tk,Rindani:2009pb} and $\gamma p$-induced processes \cite{Senol:2014naa}. Using the recent ATLAS measurement of the signal strength, $\mu^{Z\gamma}=1.3^{+0.6}_{-0.5}$ at $\sqrt{s}=13.6$ TeV \cite{ATLAS:2025aip}, the resulting upper bounds on the $CP$-violating form factor are presented in Fig. \ref{boundn}, and read
\begin{equation}
\label{boundh3}
\big|{\rm Re}\big[h^{Z\gamma}_3\big]\big|\text{, }\big|{\rm Im}\big[h^{Z\gamma}_3\big]\big|\lesssim 0.9 \text{ GeV  (95\% CL).} 
\end{equation}
This value is one order of magnitude tighter than our previous result in Ref. \cite{Hernandez-Juarez:2024iwe}, which was based on the combined analysis of the ATLAS and CMS collaborations at $\sqrt{s}=13$ TeV. Bounds consistent with those shown in Eq. \eqref{boundh3} are obtained when using the recent value for $\mu^{Z\gamma}$ reported by the CMS collaboration \cite{CMS:2026wqg}, where the data collected at center-of-mass energies of 13 and 13.6 TeV are combined.
Constraints on $CP$-violating Higgs couplings can also be derived from electric dipole moments (EDM) \cite{Chien:2015xha,Weinberg:1989dx}. In this context, bounds on the $h_3^{Z\gamma}$ form factor were obtained in effective Higgs couplings scenarios \cite{Dwivedi:2015nta, Dwivedi:2016xwm}. Using the notation in Eq. \eqref{VertexFunction}, these limits cover the range:
\begin{equation}
   10^{-2}\text{ GeV} \leq \big|h_3^{Z\gamma}\big|\leq 10^{-1}-10^{1} \text{ GeV.} 
\end{equation}
We emphasize that the upper bounds in Fig. \eqref{boundn}, attained from the measured signal strength $\mu^{Z\gamma}$ at the LHC are comparable in size to those derived from EDM observables. 

\section{Numerical analysis}\label{numsec}

In this section, we present a numerical analysis of the $CP$-violating effects
on the differential partial width in Eq.~\eqref{difpart} and on the forward--backward asymmetry. As these observables depend on the real and imaginary parts of the $CP$-conserving form factor $h_1^{Z\gamma}$, we consider its SM values given in Eq. \eqref{h1num}. We focus on charged-leptons final states.

To analyze the behavior of the differential partial width as a function of $\cos\theta$, we consider four benchmark scenarios for the real and imaginary components of $h_3^{Z\gamma}$. These scenarios are listed in Table~\ref{tabscenarios} and are chosen to be compatible with the bounds shown in Fig.~\ref{boundn}. In scenarios $I$ and $II$, we separately examine the impact of ${\rm Re}[h_3^{Z\gamma}]$ and ${\rm Im}[h_3^{Z\gamma}]$, with values close to the upper limits to maximize their contributions. In scenarios $III$ and $IV$, we consider hierarchical configurations in which one component is dominant at the level of $10^{-1}$, while the other is suppressed by one order of magnitude. To make the impact of the $CP$-odd form factor explicit, we also include the SM case for comparison. 

\begin{table}[H]
\centering
\begingroup
\setlength{\tabcolsep}{16pt}
\renewcommand{\arraystretch}{1.5}
\begin{tabular}{ccc}
Scenario & ${\rm Re}\big[h_3^{Z\gamma}\big]$ [GeV] & ${\rm Im}\big[h_3^{Z\gamma}\big]$ [GeV] \\
\hline\hline
$I$   & 0.3     & 0   \\
\hline
$II$  & 0       & 0.3 \\
\hline
$III$ & $10^{-2}$ & $10^{-1}$ \\
\hline
$IV$  & $10^{-1}$ & $10^{-2}$ \\
\hline
\end{tabular}
\caption{Different scenarios for the real and imaginary parts of the $CP$-violating form factor $h_3^{Z\gamma}$.}
\label{tabscenarios}
\endgroup
\end{table}

\begin{figure}[H]
\begin{center}
\includegraphics[width=15cm]{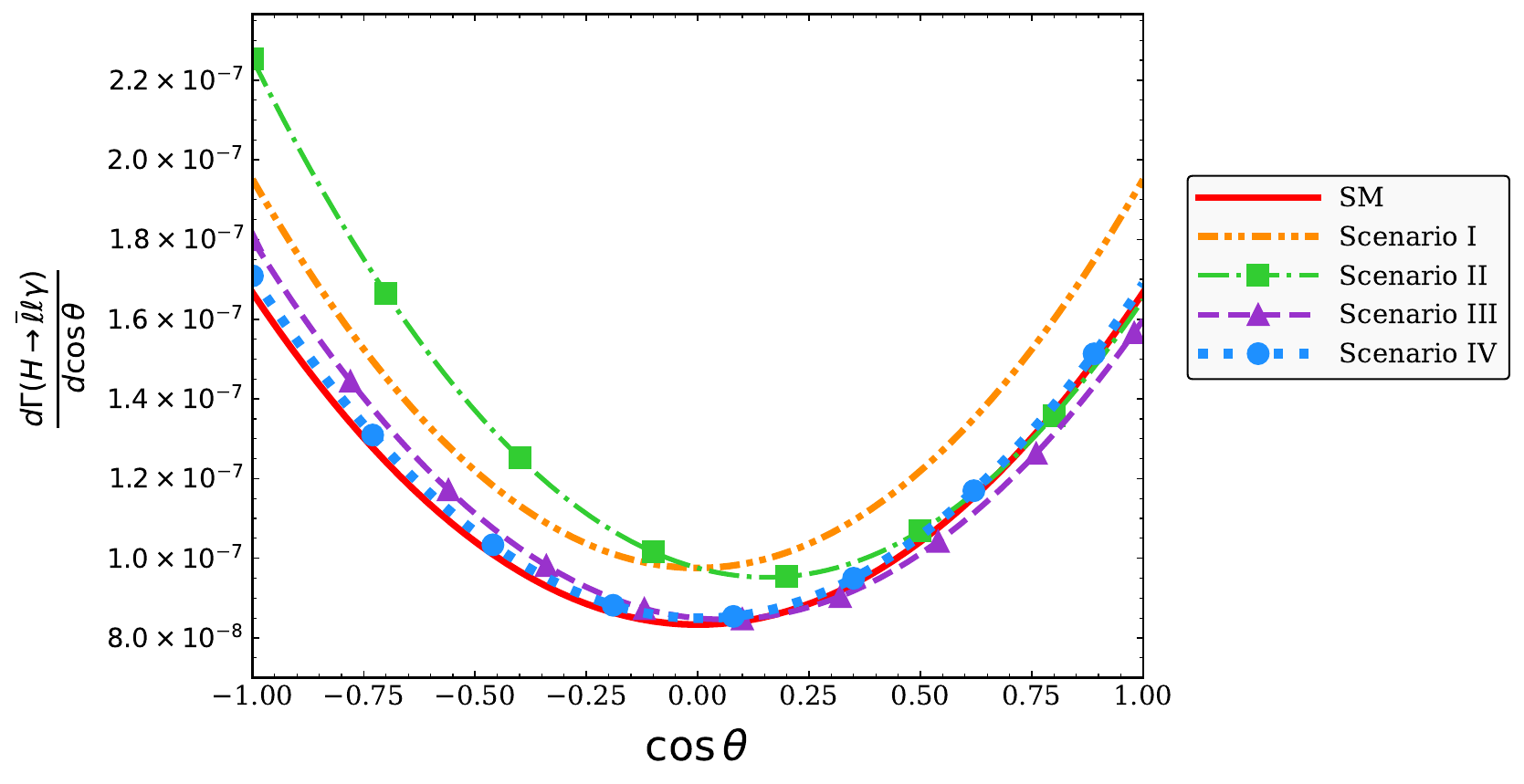}
\caption{Differential partial width of the $H\to Z\gamma\to\overline{\ell}\ell\gamma$ decay as a function of $\cos\theta$ for the SM and the benchmark scenarios defined in Table~\ref{tabscenarios}. The $CP$-conserving form factor $h_1^{Z\gamma}$ is fixed to its SM electroweak value.} \label{plotW}
\end{center}
\end{figure}

In Fig.~\ref{plotW}, we show the differential partial width of the $H\to Z\gamma\to\overline{\ell}\ell\gamma$ decay as a function of $\cos\theta$
for the SM case and the four benchmark scenarios.  For scenarios $I$ and $II$, where larger $CP$-violating effects are considered, the deviation from the SM prediction is more significant. The contribution from ${\rm Re}[h_3^{Z\gamma}]$ modifies the distribution approximately uniformly across the full angular range, whereas ${\rm Im}[h_3^{Z\gamma}]$ induces a distortion that is more pronounced for $\cos\theta\leqslant 0$. In the more realistic scenarios $III$ and $IV$, the deviations are reduced, although they remain noticeable in the case where the imaginary component dominates. The overall behavior in Fig. \ref{plotW} is consistent with previous findings reported in Ref. \cite{Anderson:2013afp}. 


\begin{figure}[H]
\begin{center}
\includegraphics[width=11cm]{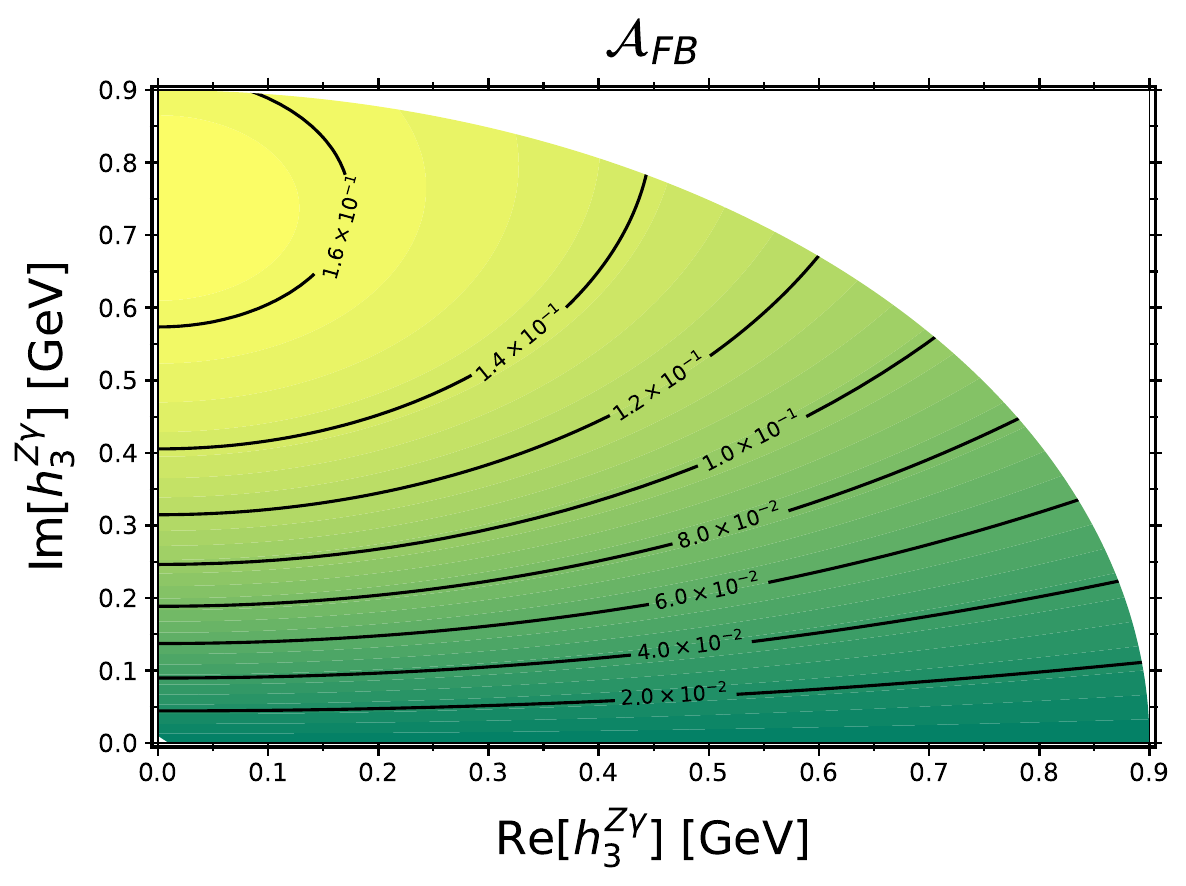}
\caption{Forward--backward asymmetry ($\mathcal{A}_{FB}$) in the allowed positive values of the real and imaginary parts of the $CP$-violating form factor $h_3^{Z\gamma}$.} \label{FBnum1}
\end{center}
\end{figure}

For the forward--backward asymmetry, we present in Fig. \ref{FBnum1} the values of $\mathcal{A}_{FB}$ in the ${\rm Re}\big[h^{Z\gamma}_3\big]$ vs ${\rm Im}\big[h^{Z\gamma}_3\big]$ plane. The analysis is restricted to the positive region consistent with the bounds shown in Fig.~\ref{boundn}, although a similar behavior is expected throughout the full parameter space. We find that for large values of the imaginary part, the asymmetry can reach $\mathcal{O}(10^{-1})$, an order of magnitude larger than that induced by  interference amplitudes \cite{Chen:2014ona}. Even for smaller contributions, with $h_3^{Z\gamma}\sim \mathcal{O}(10^{-2})$, the asymmetry can still attain appreciable values of order $10^{-2}$. In the SM, the real part of the $CP$-conserving form factor $h_1^{Z\gamma}$ is of order $10^{-1}$, while its imaginary part is about three orders of magnitude smaller. Therefore, the dominant contribution to $\mathcal{A}_{FB}$ in Eq.~\eqref{FW-exp} arises from the interference between ${\rm Re}[h_1^{Z\gamma}]$ and ${\rm Im}[h_3^{Z\gamma}]$. However, in models with heavy particles, the real and imaginary parts of $h_3^{Z\gamma}$ can reach values close to unity \cite{He:2020suf, Hernandez-Juarez:2024iwe}, without inducing significant deviations from the SM predictions for the $H\to\gamma\gamma$ and $H\to gg$ decays. In such scenarios, the interference between ${\rm Im}[h_1^{Z\gamma}]$ and ${\rm Re}[h_3^{Z\gamma}]$ may also become relevant. Nevertheless, due to the strong suppression of ${\rm Im}[h_1^{Z\gamma}]$ relative to its real part, the magnitude of the asymmetry remains predominantly controlled by ${\rm Im}[h_3^{Z\gamma}]$, as shown in Fig.~\ref{FBnum1}. Consequently, cases in which only ${\rm Re}[h_3^{Z\gamma}]$ is sizable, such as scenario $I$ in Table~\ref{tabscenarios}, are less favorable for observing $\mathcal{A}_{FB}$.

It is important to determine whether $\mathcal{A}_{FB}$ can be observed at the LHC or future colliders. To this end, we estimate the expected significance, $Z=S/\sqrt{B}$. We define $N_S$ and $N_B$ as the number of signal and background events for the process $gg\to H\to Z\gamma\to \overline{\ell}\ell\gamma$. The asymmetry signal can be interpreted as an excess in $N_S$, such that $S\sim \mathcal{A}_{FB}\, N_S$. The background includes both the $N_B$ and the $CP$-conserving contribution of the $H\to Z\gamma$ decay \cite{Chen:2014ona}, yielding
\cite{Godbole:2007cn, Dwivedi:2016xwm}
\begin{equation}
\label{sig1}
Z=\mathcal{A}_{FB}\frac{N_S}{\sqrt{N_S+N_B}},
\end{equation}
The ATLAS collaboration has recently reported an observed significance of 2.5 standard deviations for the $\overline{\ell}\ell\gamma$ final state \cite{ATLAS:2025aip}, combining Run-2 and Run-3 data with 140 fb$^{-1}$ and 165 fb$^{-1}$ luminosities, respectively. Since the irreducible and reducible backgrounds dominate \cite{PhysRevD.86.033010, Chen:2017plj}, we approximate $\sqrt{N_S+N_B}\approx\sqrt{N_B}$. Thus, the relation between the reported ATLAS significance and Eq. \eqref{sig1} is given by $S\sim 2.5\, \mathcal{A}_{FB}$. Rescaling with luminosity, we obtain
\begin{equation}\label{sigfin}
Z = \frac{\mathcal{A}_{FB}}{(0.12)}\sqrt{\frac{\mathcal{L}}{3000\text{ fb}^{-1}}}.
\end{equation}
This expression agrees with Refs \cite{Chen:2014ona, Chen:2017plj}, where the signal and backgrounds have been simulated. Eq. \eqref{sigfin} indicates that for the high-luminosity LHC (HL-LHC) with $\mathcal{L}=3000$ fb$^{-1}$ and values of ${\rm Im}\big[h^{Z\gamma}_3\big]\sim\mathcal{O}(10^{-1})$, the $\mathcal{A}_{FB}$ asymmetry could reach the discovery level. The observation of the forward--backward asymmetry would be clear evidence of $CP$ violation.

As pointed out above, the imaginary part of $h_3^{Z\gamma}$ provides the dominant contribution to the forward--backward asymmetry. The values of ${\rm Im}[h_3^{Z\gamma}]$ considered so far have been derived solely from experimental constraints. Nevertheless, an absorptive part may arise from loop-induced processes when the particles circulating in the loop can go on shell. At the one-loop level, this possibility can occur, for instance, in the presence of FCNC couplings of the Higgs and $Z$ bosons. Therefore, the magnitude of ${\rm Im}[h_3^{Z\gamma}]$ is model dependent. In Ref.~\cite{Hernandez-Juarez:2024iwe}, FCNC couplings involving the top quark were found to generate ${\rm Re}[h_3^{Z\gamma}]$ of order $10^{-5}$. Although an absorptive part can arise for lighter particles, their FCNC contributions are expected to be smaller than those associated with the top quark. Under this assumption, ${\rm Im}[h_3^{Z\gamma}]$ would be strongly suppressed, making $\mathcal{A}_{FB}$ challenging to observe at the LHC.


\section{Conclusions}\label{secconclu}

In this work, we analyzed the forward--backward asymmetry in the $H\to Z\gamma\to \overline{f}f\gamma$ decay in the context of recent LHC results. This asymmetry is induced by the presence of a $CP$-violating form factor and imaginary parts of the $HZ\gamma$ vertex. Although $\mathcal{A}_{FB}$ has been explored in previous studies, we provided its explicit analytical expression for the first time. Furthermore, we studied angular distributions, which are sensitive to $CP$-violating effects. To estimate the magnitude of the asymmetry, we derived bounds on the real and imaginary components of the $CP$-violating form factor $h_3^{Z\gamma}$ using current LHC results. We obtained an upper limit of about $0.9$ GeV, comparable to those from EDM constraints. We found that the asymmetry can reach values of order $10^{-1}$ for large ${\rm Im}[h_3^{Z\gamma}]$. Under HL-LHC conditions, $\mathcal{A}_{FB}$ could become experimentally accessible at the discovery level. The observation of this forward--backward asymmetry would indicate the existence of new sources of $CP$ violation, which are necessary to explain the matter-antimatter asymmetry.

\section{Acknowledgments}
This work was supported by Sistema Nacional de Investigadores (Mexico).  A.I. Hern\'andez-Ju\'arez acknowledges additional support from the program "Estancias Posdoctorales por M\'exico". We thank Arely Cort\'es-Gonz\'alez for valuable discussions and I. Garc\'ia-M\'arquez for assistance in the preparation of Fig. \ref{plano}.

\appendix
\section{Kinematics}\label{kinematicappendix}

\begin{figure}[H]
\begin{center}
\includegraphics[width=9cm]{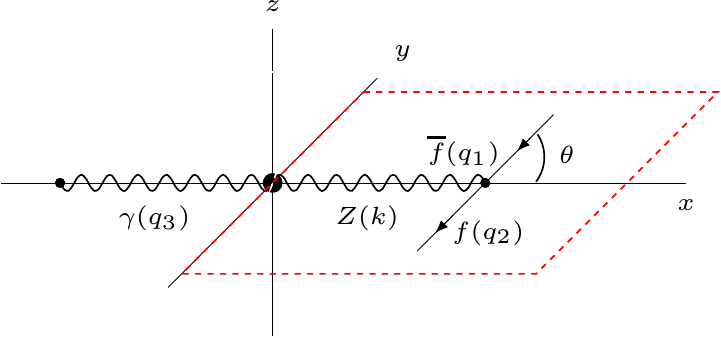}
\caption{Kinematic configuration and momentum assignments for the $H\to\overline{f}f\gamma$ decay.} \label{plano}
\end{center}
\end{figure}

This appendix presents the kinematics employed in the computation of the $H\to Z\gamma\to\overline{f}f\gamma$ amplitude. We follow the momentum assignments shown in Fig.~\ref{plano}.

\subsection{Higgs boson rest frame}
In the Higgs rest frame, the four-momenta of the Higgs boson, the $Z$ boson, and the photon can be expressed as follows
\begin{align}
    &q^\mu=(m_H, 0),\\
    &k^\mu=(\frac{m_H^2+m_Z^2}{2m_H},\vec{k})\\
    &q_3^\mu=(\frac{m_H^2-m_Z^2}{2m_H},-\vec{k}),
\end{align}
with the magnitude of the three-momentum $\vec{k}$:
\begin{equation}
    \|\vec{k}\|=\frac{m_H^2-m_Z^2}{2m_H}.
\end{equation}

\subsection{$Z$ gauge boson rest frame}

In the rest frame of the $Z$ boson, the corresponding four-momenta are expressed as
\begin{align}
    &k^\mu=(m_Z,0),\\
    &q_1^\mu=(m_Z/2,\vec{q}),\\
    &q_2^\mu=(m_Z/2,-\vec{q}),
\end{align}
where, according to the kinematical configuration shown in Fig.~\ref{plano}, the three-momentum $\vec{q}$ is given by
\begin{equation}
    \vec{q}=\frac{\sqrt{m_Z^2-4m_\ell^2}}{2}\ (\cos\theta,\sin\theta,0).
\end{equation}

\subsection{Lorentz invariants}\label{lorinvcal}
In the evaluation of the squared amplitude for the $H\to Z\gamma\to\overline{\ell}\ell\gamma$ decay, scalar products between four-momenta defined in different reference frames are required. In the Higgs rest frame, the relevant invariants are given by
\begin{align}
    &q_1\cdot q_3=\frac{(m_H^2-m_Z^2)}{4}(1+\cos\theta),\\
    &q_2\cdot q_3=\frac{(m_H^2-m_Z^2)}{4}(1-\cos\theta),
\end{align}
which are obtained by performing a Lorentz transformation between frames with velocity $\nu=\frac{m_H^2-m_Z^2}{m_H^2+m_Z^2}$.

\subsection{ Phase space}\label{spacephaseapp}
The phase space ($dR_3$) for the $H\to Z\gamma\to \overline{f}f\gamma$ decay can be expressed as
\begin{equation}
    dR_3=\frac{d^3\vec{q}_1}{(2\pi)^3 2E_{\vec{q}_1}}\frac{d^3\vec{q}_2}{(2\pi)^3 2E_{\vec{q}_2}}\frac{d^3\vec{q}_3}{(2\pi)^3 2E_{\vec{q}_3}}(2\pi)^4\delta^4(q-q_1-q_2-q_3).
\end{equation}
Following the approach in Refs. \cite{Cheng:1993ah,Cabibbo:1965zzb}, we introduce the identity
\begin{equation}
    \int \frac{d^3\vec{k}}{2E_{\vec{k}}}dK\ \delta^4(k-q_1-q_2)=1,
\end{equation}
where $K=k^2$ and $E_{\vec{k}}=\sqrt{|\vec{k}|^2+K}$. The phase space can then be written as
\begin{equation}
    dR_3=\frac{dK}{(2\pi)^5}I_q I_k,
\end{equation}
where $I_q$ and $I_k$ are evaluated in the Higgs and $Z$ boson rest frames, respectively, and are given by
\begin{align}
    I_q&=\int \frac{d^3\vec{k}}{2E_{\vec{k}}} \frac{d^3\vec{p}_3}{2E_{\vec{p}_3}}\delta^4(q-k-p_3)\nonumber\\
    &=\frac{\pi}{2m_H^2}(m_H^2-m_Z^2),
\end{align}

\begin{align}
    I_k&=\int \frac{d^3\vec{q}_1}{2E_{\vec{q}_1}} \frac{d^3\vec{q}_2}{2E_{\vec{q}_2}}\delta^4(k-q_1-q_2)\nonumber\\
    &=\frac{\pi}{4}d\cos\theta.
\end{align}

\bibliography{Biblio}

\end{document}